\documentclass[
draft            % use draft while you are working on the paper
  ]
  {aipproc}

  \layoutstyle{6x9}

\begin{document}

\title{Modern optical potentials and the role of nuclear structure}


\author{S. Karataglidis}{
  address={School of Physics, University of Melbourne, Victoria, 3010,
  Australia}
}

\begin{abstract}
Microscopic  descriptions of  exotic nuclei  are the  subject  of much
experimental and  theoretical effort. Not  only are such  important in
their own  right but  are also necessary  for applications  in nuclear
astrophysics.  Evaluations of model  wave functions  may be  done with
analyses  of elastic  and  inelastic scattering  from hydrogen.  Those
require a realistic model  of nucleon-nucleus scattering as scattering
from  hydrogen   translates  to  proton  scattering   in  the  inverse
kinematics. The Melbourne $g$-folding model for intermediate energy is
presented along  with various examples. Implications  for existing and
future experimental and theoretical work are discussed.
\end{abstract}

\maketitle

%%%%%%%%%%%%%%%%%%%%%%%%%%%%%%%%%%%%%%%%%%%%
%% MAINMATTER
%%%%%%%%%%%%%%%%%%%%%%%%%%%%%%%%%%%%%%%%%%%%
\section{Introduction}

The microscopic  facets of the  structures of exotic nuclei  have been
the   subject  of  increased   experimental  and   theoretical  study,
particularly with data now available  for the scattering of heavy ions
from hydrogen at intermediate energies. Such data are complementary to
those obtained for breakup  reactions, which only probe the asymptotic
part  of  the  initial  state  wave function  of  the  exotic  nucleus
\cite{Ha96,Es96}. Also,  it has been  established that the  breakup of
$^6$He  is  a  two-step  process  \cite{Al98}, with  the  final  state
interactions greatly influencing  the reaction process. To investigate
the wave functions of exotics  at a deeper level one requires analyses
of complementary scattering data.

In the absence of electron scattering data, intermediate-energy proton
scattering represents the best  probe of the microscopic structures of
exotic nuclei.  This was  illustrated by Karataglidis  \textit{et al.}
who compared electron and proton elastic and inelastic scattering from
$^{12}$C \cite{Ka95} and $^{6,7}$Li  \cite{Ka97a}. They found that the
behaviour  with momentum transfer  of the  form factors  from electron
scattering  was found  also in  the differential  cross  sections from
intermediate energy proton  scattering, reflecting the deficiencies in
the wave  functions from the underlying assumed  model structure. Such
comparisons were only possible  when credible models of scattering for
both electrons and protons from nuclei were specified.

It is well-known that few-body descriptions of exotics, especially the
halo nuclei,  are able to  describe breakup reactions as  those models
are  able  to  give  the  correct asymptotic  behaviour  of  the  wave
functions. A problem, however, exists when attempting to use such wave
functions  in descriptions  of  scattering from  hydrogen: a  credible
description of the  structure of the core is  necessary to account for
the  full density,  which is  required to  analyse  nucleon scattering
data.  That  was found  in  analyses  of  scattering from  $^9$Li  and
$^{11}$Li  \cite{Cr96}.  More   recently,  such  models  incorporating
multiple   scattering  expansions   of   the  nucleon-nucleon   ($NN$)
scattering amplitudes \cite{Cr99,Cr02}  have sought to describe proton
scattering from halos using few-body models. However, predictions from
those  models are  still  prone to  significant  changes arising  from
problems  in  specifying  the  density  of the  core  \cite{Cr99}.  At
intermediate  energies,  medium  modifications  are essential  in  the
specification  of the $NA$  optical potential  (OMP) and  the multiple
scattering expansions,  taken to second order \cite{Cr99},  are only a
gross approximation.  A full  $g$-matrix specification of  the optical
potential is needed to account for those corrections \cite{Am00}.

As microscopic models of the nucleon-nucleus ($NA$) optical potentials
are based on effective  $NN$ interactions, these problems are overcome
when models  of structure which  admit nucleon degrees of  freedom are
used.  Herein,  a  description  of  the  Melbourne  optical  potential
\cite{Am00} which accounts for  scattering from both stable and exotic
nuclei  self-consistently  is  presented.  That model  has  been  used
successfully  with the  shell  model \cite{Am00},  Skyrme-Hartree-Fock
(SHF) models \cite{Ka02}, and  the RPA \cite{Du05,Du05a}. After giving
a brief  description of  the model, various  results are given  with a
view  to future  experiments which  may  be done  at radioactive  beam
facilities. Concluding remarks follow.

\section{The $g$-folding optical model - Melbourne force}
The Melbourne force is discussed  in detail in a recent review article
by Amos \textit{et al.} \cite{Am00},  to which the reader is referred.
A brief overview of the model is presented herein with emphasis on how
structure  enters  into  the  description  of  elastic  and  inelastic
scattering.

The  optical potential  for  $NA$ scattering  is  associated with  the
elastic   scattering  channel.   Following   the  Feshbach   formalism
\cite{Fe62}, we  split the Hilbert  space into the  elastic scattering
channel  ($P$  space)  and   non-elastic  channels  ($Q$  space).  The
Schr\"odinger equation then becomes,  with $P$ and $Q$ projectors onto
the respective spaces
\begin{eqnarray}
  \left( E - H_{PP} \right) \left| \Psi^{(+)} \right\rangle & = &
  H_{PQ}\left| \Psi^{(+)} \right\rangle \nonumber \\
  \left( E - H_{QQ} \right) \left| \Psi^{(+)} \right\rangle & = &
  H_{QP} \left| \Psi^{(+)} \right\rangle\; ,
\end{eqnarray}
where   $H_{XY}  =   XHY$.   Recoupling,  and   taking  the   one-body
approximation gives the Schr\"odinger equation for the projectile wave
function, \textit{viz.}
\begin{equation}
  \left\{ E - H_0 - \left\langle \Phi_{gs} \left| V \right| \Phi_{gs}
  \right\rangle - \left\langle \Phi_{gs} \left| V G^{(+)}_{QQ} V
  \right| \Phi_{gs} \right\rangle \right\} \left| \chi^+ \right\rangle
  = 0
\end{equation}
where $G^{(+)}_{QQ} = \left[ E - H_{QQ} + i\varepsilon \right]^{-1}$,
from which the optical potential is defined as
\begin{equation}
  U = \left\langle \Phi_{gs} \left| V \right| \Phi_{gs} \right\rangle
  + \left\langle \Phi_{gs} \left| V G^{(+)}_{QQ} V \right| \Phi_{gs}
  \right\rangle \; .
\end{equation}

Specification  of the optical  potential is  a many-body  problem with
explicit dependence  on the target  ground state wave function.  It is
complex,  nonlocal   and  energy  dependent,  through   $V$  and  also
$G^{(+)}_{QQ}$. The  second term  is the Dynamic  Polarising Potential
(DPP)  and defines  how coupling  to nonelastic  channels  varies with
energy.  Specifically, such  coupling may  be cast  into  three energy
regimes:
\begin{description}
  \item[Low energy] For $E  < 10$~MeV, explicit coupling to specified,
  discrete  low-lying excited  states of  the target  is  necessary. A
  recent  development  in  the  construction  of such  models  is  the
  Multi-Channel Algebraic Scattering  theory \cite{Am03,Ca05} which is
  the subject of another presentation at this meeting \cite{Am05};
  \item[Giant resonances] Between 10  and 25~MeV coupling to the giant
  resonances  becomes important  \cite{Ge75}. One  important exception
  are the He isotopes, for which there are no giant resonances;
  \item[Intermediate and High energies]  At higher energies and as the
  level  density  becomes high,  coupling  to  excited  states may  be
  handled implicitly  by using  folding models based  on the  $NN$ $g$
  matrices for infinite matter.
\end{description}

An example of the last is the Melbourne $g$-folding model \cite{Am00}.
The model  takes as its  basis an effective $NN$  interaction obtained
from the  $g$ matrices of the  bare $NN$ interaction.  For an incident
nucleon, momentum $\mathbf{p}_0$ in  collision with a nucleon embedded
in infinite  matter with  momentum $\mathbf{p}_1$, those  $g$ matrices
are  solutions of  the Bruckner-Bethe-Goldstone  equation  in momentum
space, \textit{viz.}
\begin{equation}
  g( \mathbf{q}, \mathbf{q}'; \mathbf{K} ) = V(
  \mathbf{q}, \mathbf{q}') + \int V( \mathbf{q}',
  \mathbf{k}' ) \frac{ Q( \mathbf{k}', \mathbf{K}; k_f) }{ \left[ E(
  \mathbf{k}, \mathbf{K} ) - E( \mathbf{k}',
  \mathbf{K} ) \right] } g( \mathbf{k}', \mathbf{q}; \mathbf{K})
  \; d\mathbf{k}' \, ,
\end{equation}
where  $\mathbf{k} =  (\mathbf{p}_0 -  \mathbf{p}_1)$ is  the relative
momentum  and  $\mathbf{K}$  is  the  centre-of-momentum  of  the  two
particles.  Primes   denote  the  equivalent  set   of  momenta  after
scattering. $Q$ is  a Pauli-blocking operator and the  energies $E$ in
the propagator contain auxiliary potentials which model the effects of
the nuclear medium \cite{Ha70}. As  $Q$ and energies $E$ are dependent
on   $\mathbf{k}'$,   in  practice   they   are   replaced  by   their
angle-averaged values. This has been  shown to be a good approximation
for nuclear  densities above $\sim  15\%$ \cite{Le78,Ch89}, and  is an
important  consideration  for  scattering  from  exotic  nuclei  where
scattering is  observed as from  the core in  the case of  halo nuclei
\cite{Ka00}.

Once the $g$ matrices in infinite matter are obtained, they are mapped
to those for finite nuclei  in coordinate space \cite{Am00} by folding
in  the  specified (model)  density  of  the  target. The  mapping  to
coordinate space is achieved by means of a double Bessel transform and
allows for the explicit specification of central, tensor, and two-body
spin-orbit  terms  as  sums  of  Yukawa functions.  (This  is  also  a
practical consideration: the DWBA  suite of programs \cite{Ra98} which
are  used   to  calculate  observables  require   a  coordinate  space
representation of  the potential.)  Once these effective  $g$ matrices
($g_{eff}$)  have  been  obtained,  the  nonlocal,  complex,  OMP  for
scattering is defined as
\begin{eqnarray}
  U( \mathbf{r}, \mathbf{r}'; E ) & = &
  \delta(\mathbf{r} - \mathbf{r}') \sum_i n_i \int
  \varphi^{\ast}_i( \mathbf{s} ) g_D( \mathbf{r}, \mathbf{s}; E)
  \varphi_i(\mathbf{s}) \, d\mathbf{s} \nonumber \\
  & & + \sum_i n_i \varphi_i(\mathbf{r}') g_E( \mathbf{r},
  \mathbf{r}'; E ) \varphi_i \nonumber \\
  & = & U_D( \mathbf{r}, E ) \delta( \mathbf{r} -
  \mathbf{r}') + U_E( \mathbf{r}, \mathbf{r}'; E )
  \label{OMP}
\end{eqnarray}
where  $D$  and  $E$ denote  the  direct  and  exchange terms  of  the
effective  interaction  respectively.  Nuclear  structure  information
enters via the occupation numbers $n_i$ for each orbit $i$. The direct
term is  the well-known $g\rho$ form  of the optical  potential and is
local.  The  nonlocality  arises  from the  explicit  exchange  terms;
neglecting  such terms  can lead  to serious  problems  \cite{De00}. A
credible model of  structure is necessary in the  specification of the
optical potential.

The  single particle  wave functions  entering in  Eq.~(\ref{OMP}) are
usually  assumed to  be of  harmonic  oscillator (HO)  form. For  most
nuclei  this is  a reasonable  assumption and  is consistent  with the
underlying  shell   model.  However,  for  halo  nuclei   it  is  more
appropriate to  use Woods-Saxon  (WS) wave functions  \cite{Ka00} with
binding  energies  of the  orbits  occupied by  the  halo  set to  the
separation energy of the single nucleon in the halo.

Inelastic   scattering   may  be   calculated   in  a   distorted-wave
approximation (DWA) with the  $g_{eff}$ as the operators effecting the
transition. The transition amplitude may  be written, with '0' and '1'
denoting the projectile and bound state nucleon, respectively, as
\begin{equation}
  T^{M_f M_i \nu' \nu}_{J_f J_i}( \theta ) = \left\langle
  \chi^{(-)}_{\nu'}(0) \right| \left\langle \Psi_{J_f M_f} \left| A
  g_{\mathrm{eff}}(0,1) \mathcal{A}_{01} \left\{ \right| \Psi_{J_i M_i}
  \right\rangle \left| \chi^{(+)}_\nu(0) \right\rangle \right\}\; .
  \label{inelastic}
\end{equation}
In  Eq.~(\ref{inelastic}),   the  distorted  wave   function  for  the
projectile  is  denoted  by  $\chi$,  and  $\mathcal{A}_{01}$  is  the
antisymmetrization  operator  for   the  projectile  and  bound  state
nucleon.  For a  spin-zero  target, one  obtains  after expanding  the
many-body wave function
\begin{eqnarray}
  T^{M_f M_i \nu' \nu}_{J_f J_i}( \theta ) & = & \sum_{\alpha_1 m_1
  \alpha_2 m_2} \frac{ (-1)^{j_1 - m_1} }{ \sqrt{ 2J_f + 1 } }
  \left\langle j_2 \, m_2 \, j_1 \, -m_1 \right. \left| J_f \, M_f
  \right\rangle \left\langle J_f \left\| \left[ a^{\dag}_{\alpha_2}
  \times \tilde{a}_{\alpha_1} \right]^{J_f} \right\| 0 \right\rangle
  \nonumber \\
  & \times & \left\langle \chi^{(-)}_{\nu'}(0) \right| \left\langle
  \varphi_{\alpha_2}(1) \left| A g_{\mathrm{eff}}(0,1)
  \mathcal{A}_{01} \left\{ \right| \varphi_{\alpha_1}(1) \right\rangle
  \left| \chi^{(+)}_\nu(0) \right\rangle \right\} \; ,
\end{eqnarray}
where $\alpha = \{ l,m,j \}$. 

\section{Nuclear Structure Facets}

The optical  model potential  in the $g$-folding  model is  a one-body
operator  with respect  to the  target  (bound) nucleons,  and so  one
requires  specification  of   the  one-body  density  matrix  elements
(OBDME), \textit{viz.}
\begin{equation}
  S_{\alpha_1 \alpha_2 J} = \left\langle J_f \left\| \left[
  a^{\dag}_{\alpha_2} \times \tilde{a}_{\alpha_1} \right]^J \right\|
  J_i \right\rangle \;.
\end{equation}
Various models  have been utilised but,  for the most  part, the shell
model  has  been  used  to  specify  the  OBDME.  Others  include  the
Skyrme-Hartree-Fock (SHF) and the RPA,  results from both of which are
presented below. Note that in specifying  the OMP one must keep to the
level of  the density matrix  \textit{elements} as that  preserves the
nonlocality. Use  of the density itself  requires gross approximations
to be made in the handling of the nonlocal exchange terms. That may be
problematic \cite{De00}.

\section{Results}
The BonnB  $NN$ interaction was used  to obtain the  $g_{eff}$ for all
results  presented  herein.  All  results  were  obtain  using  DWBA98
\cite{Ra98}  from single-shot  calculations; there  was no  fitting to
data. The review \cite{Am00} present  most results obtained up to that
time, and includes a discussion on the connection between electron and
proton scattering. A  subset of those results are  discussed below, as
well as those obtained since the review.

\begin{figure}
  \includegraphics*[height=0.3\textheight]{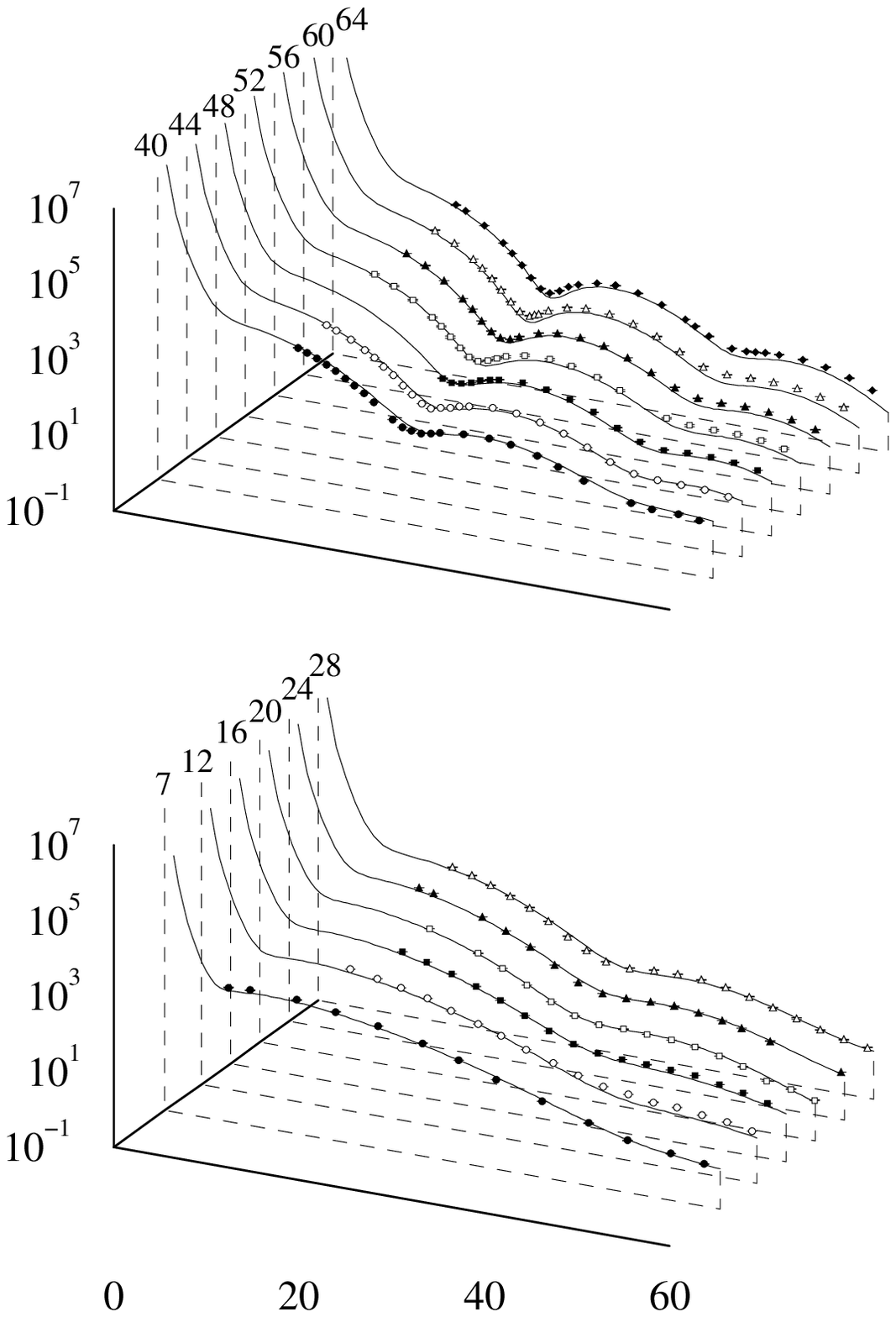}
  \includegraphics*[height=0.3\textheight]{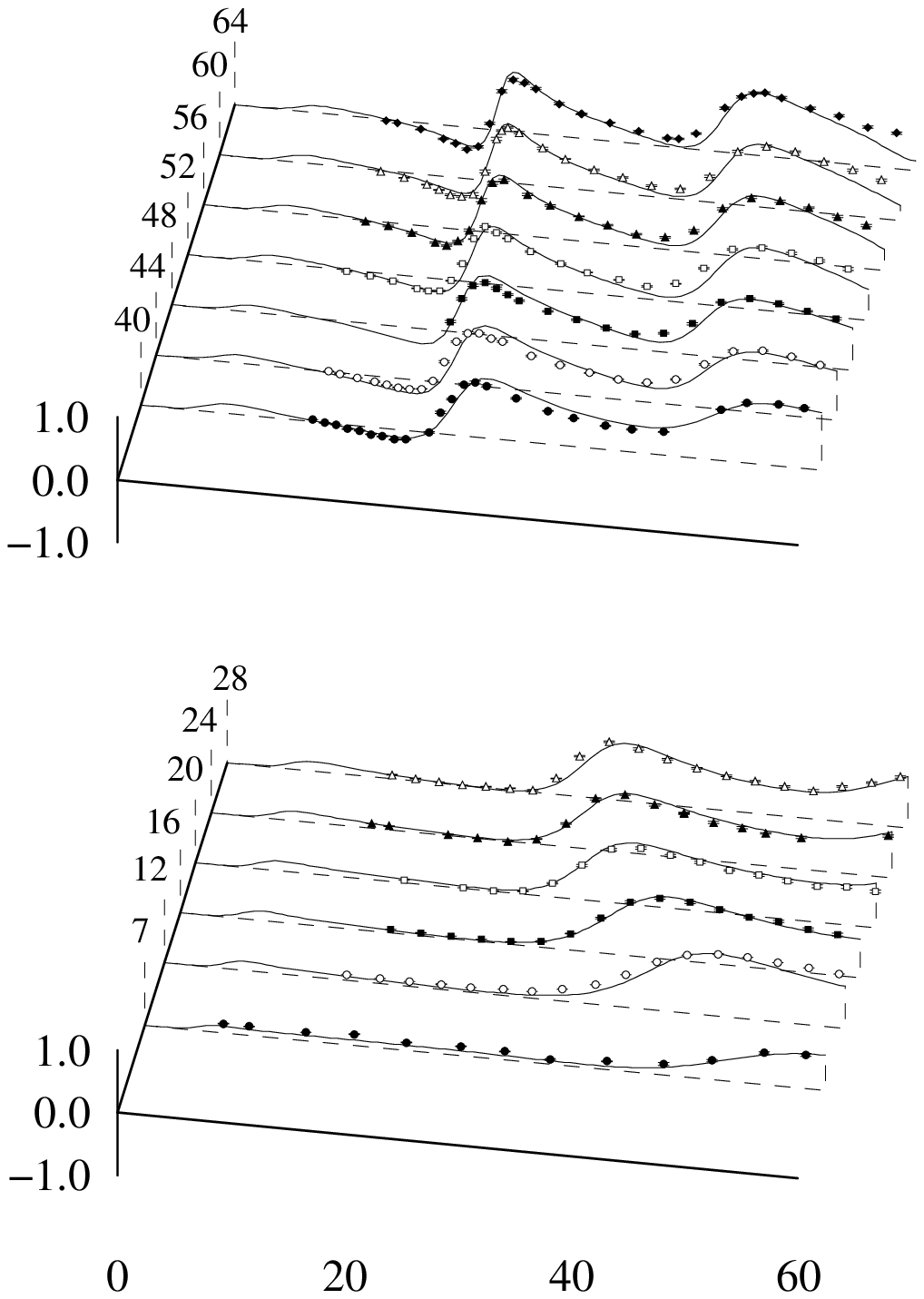}
  \caption{\label{mass65} Differential cross sections (left) and
  analysing powers (right) for the elastic scattering of 65~MeV
  protons from various nuclei to mass 64.}
\end{figure}

Systematic analyses  of elastic scattering across the  mass range, and
for  several energies,  have  been reported  (see, eg.,  \cite{Am00}).
Results of analyses of  elastic scattering differential cross sections
and analysing powers for the  scattering of 65~MeV protons from nuclei
up to mass-64 are presented in Fig.~\ref{mass65}.
Clearly the differential cross sections and analysing powers at 65~MeV
are well reproduced by the  model. Of particular note is the excellent
reproduction of the observables'  dependence with momentum transfer as
one increases the mass.

Fig.~2  displays  the  differential  cross  section  for  the  elastic
scattering  of  $^6$He  ions  from  hydrogen  at  $41A$~MeV  (data  of
Lagoyannis  \textit{et  al.} \cite{La01})  as  well  as the  inelastic
scattering cross section to the  $2^+$ state. The use of WS functions,
depicted  by the  solid  line, to  specify  the density  of $^6$He  as
consistent with a neutron halo gives better agreement with the elastic
scattering  data. (The  results using  HO functions  are given  by the
dashed lines.) At these energies, the proton does not observe the halo
directly but rather its effect in depleting the neutron density in the
core  to  the halo.  The  effect  is observed  as  a  decrease in  the
differential cross  section at large  angles. As $E2$  transitions are
surface-peaked,  the  halo  is  better illustrated  in  the  inelastic
scattering to the  $2^+$ state as an enhancement  in the cross section
around $30^{\circ}$.

The SHF model has been used to describe systematic behaviour in exotic
nuclei as  one approaches  the drip lines.  One may evaluate  the wave
functions  obtained therefrom  in analyses  of reaction  cross section
data. We  compare the results of  calculations made at  65 and 200~MeV
using SHF  wave functions  with the SkX*  force \cite{Ba00}  in Fig.~3
with estimates by Carlson \cite{Ca96}. The level of agreement is quite
good and  the calculations  exhibit the Coulomb  shift, which  is more
pronounced at 65~MeV. Of note  is that as one increases neutron number
one moves away from the line of minimal isospin. \\
\begin{minipage}[htb]{0.4\linewidth}
  \centering\includegraphics*[height=0.25\textheight]{he6_41.eps}\\
  {\small  \textbf{FIGURE  2.}  Differential  cross sections  for  the
  elastic and inelastic scattering of $^6$He from hydrogen.}
\end{minipage}\hfill
\begin{minipage}[htb]{0.4\linewidth}
  \centering\includegraphics*[height=0.25\textheight]{react.eps}\\
  {\small \textbf{FIGURE 3.} Reaction cross sections for S, Ar, and Ca
  isotopes at 65 and 200 MeV.}
\end{minipage}\\ \\
This illustrates a possible signature for exotic structures.

Fig.~\ref{sn_density} shows  the proton and neutron  densities for the
\setcounter{figure}{3}
\begin{figure}[h]
  \includegraphics*[height=0.15\textheight]{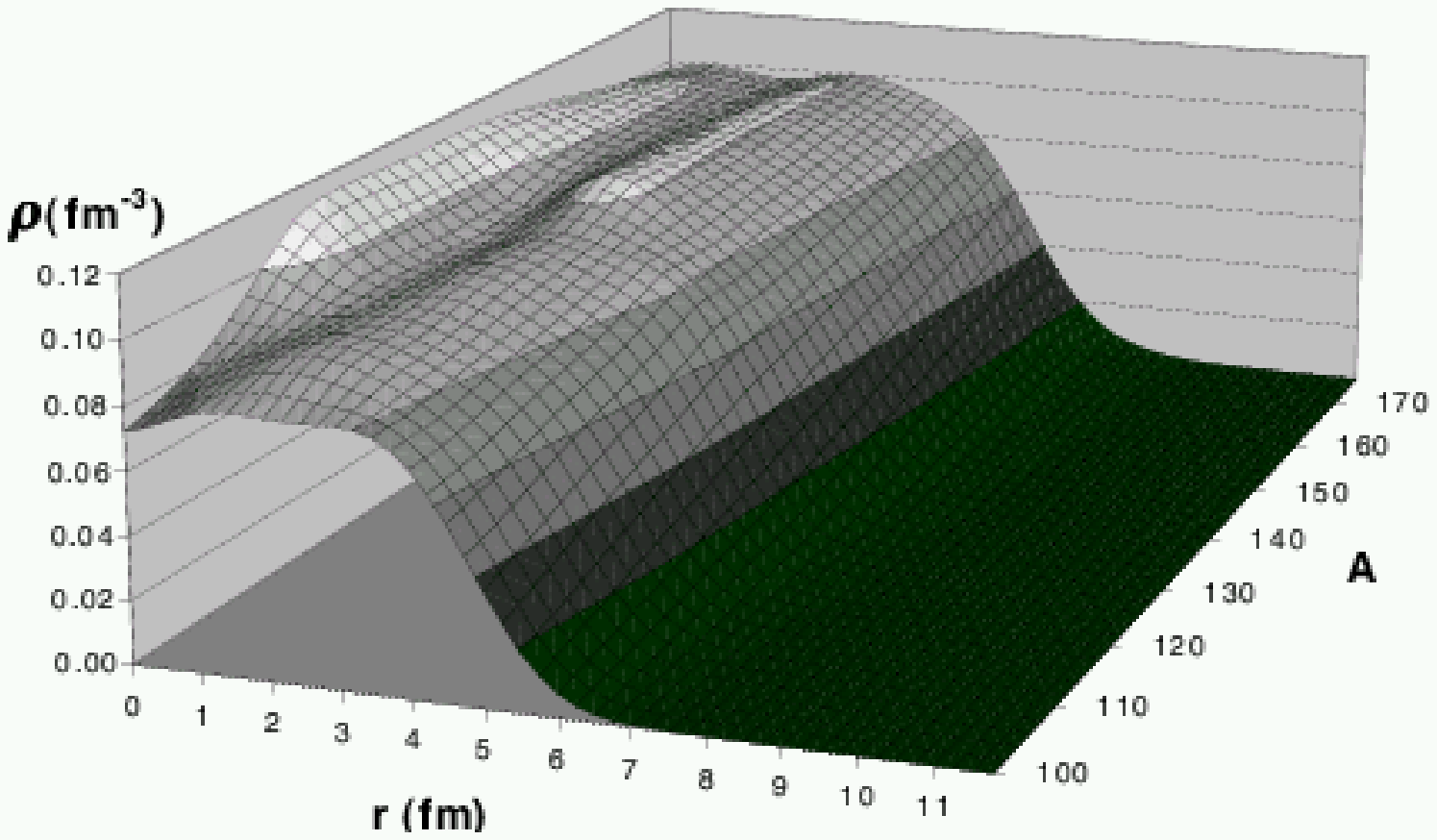}
  \includegraphics*[height=0.15\textheight]{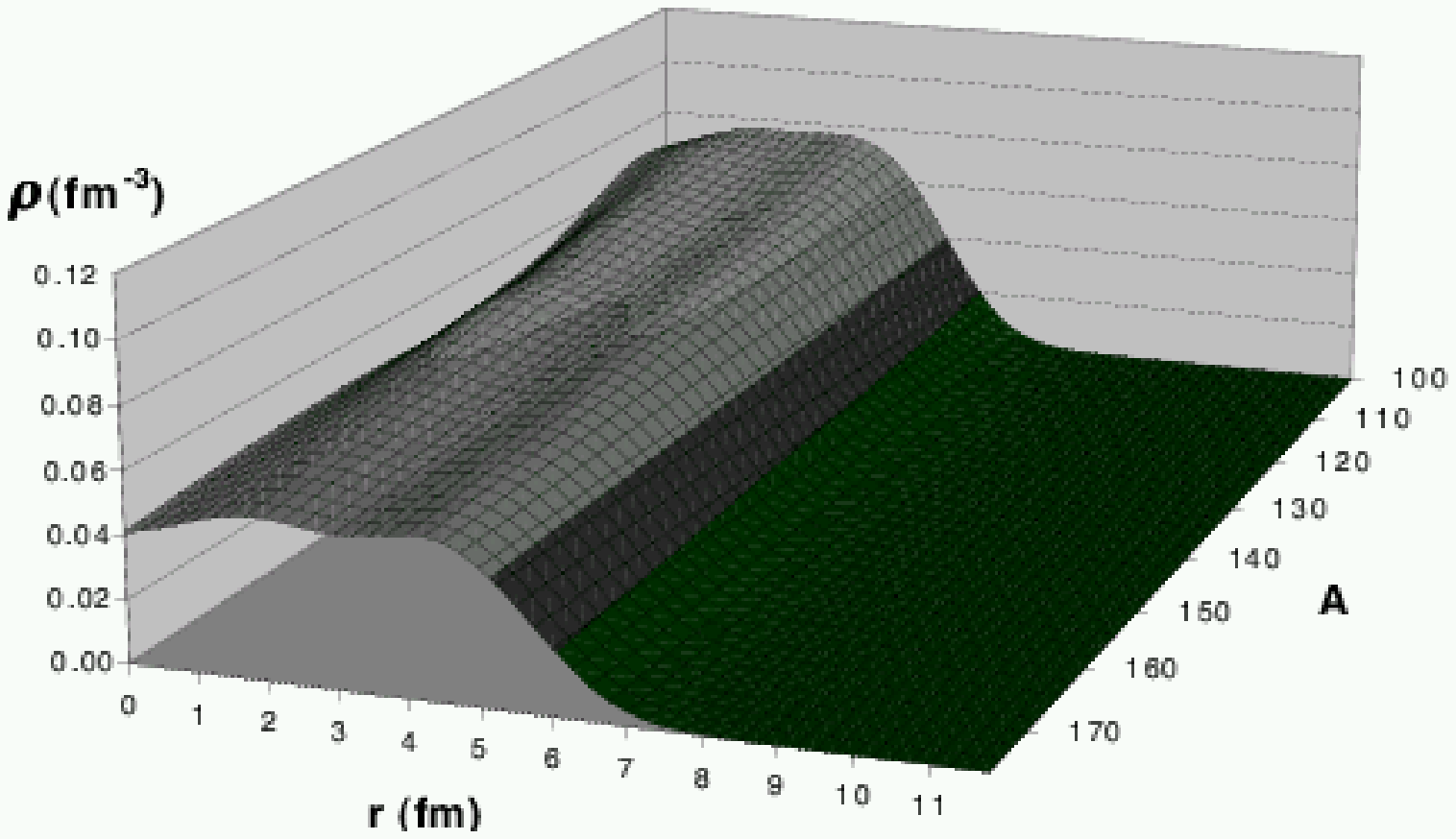}
  \caption{\label{sn_density}   Neutron  (left)  and   proton  (right)
    densities for the Sn isotopes from the SHF model.}
\end{figure}
Sn  isotopes from  $^{100}$Sn to  $^{168}$Sn as  obtained from  an SHF
model  using  the  SLy4  force \cite{Am04}.  Those  densities  exhibit
various  structures as one  increases mass,  notably an  indication of
neutron   halos   around   $A=150$.  Fig.~\ref{sn_cross}   shows   the
differential  cross sections  and  polarisations for  a  sample of  Sn
isotopes as compared to data at 40~MeV.
\begin{figure}[h]
  \includegraphics*[height=0.27\textheight]{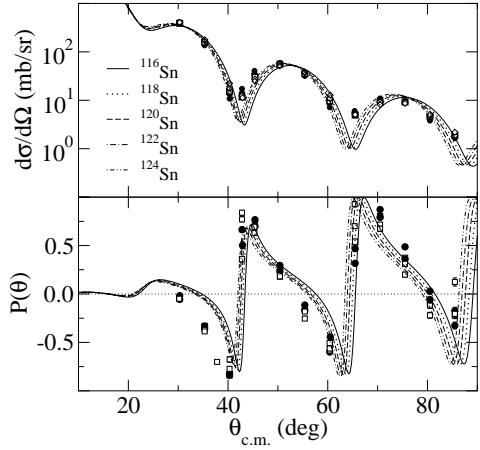}
  \caption{\label{sn_cross} Differential cross sections and
    polarisations for 40~MeV proton scattering from a sample of Sn
    isotopes.}
\end{figure}
The level of agreement between  the results and data is reasonable and
extends naturally to the variation with mass.

\begin{figure}
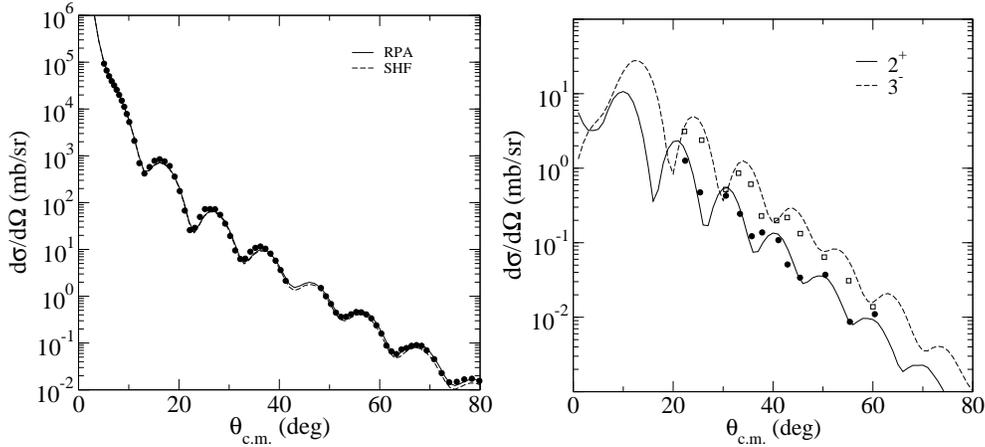

  \includegraphics*[height=0.27\textheight]{pb_fig1.eps}
  \includegraphics*[height=0.26\textheight]{pb_fig2.eps}
  \caption{\label{lead} Differential cross section for 121~MeV elastic
  (left)  and   135~MeV  inelastic  (right)   proton  scattering  from
  $^{208}$Pb.}
\end{figure}

Fig.~\ref{lead} displays the results  for elastic and inelastic proton
scattering   from  $^{208}$Pb  \cite{Du05a},   at  121   and  135~MeV,
respectively.
The  RPA model  was used  to obtain  the ground  state  and transition
densities  as  required. The  elastic  scattering  cross section  data
\cite{Na81} are  also compared  to the results  of an  SHF calculation
using the SkM* force  \cite{Ka02}. The agreement with data illustrates
that both the  SHF and RPA models for $^{208}$Pb  for the ground state
are reasonable. Of the two models, only the RPA may be able to specify
transitions densities and  those have been used to  obtain the results
of  the inelastic  scattering. Data  \cite{Ad80} and  results  for the
transitions to  the $2^+$ and $3^-$  states are compared  in the right
panel of Fig.~\ref{lead}. Use of  the RPA allows for a self-consistent
analysis  of  data  leading to  the  even  and  odd parity  states  in
$^{208}$Pb. That both sets of results are out of phase with respect to
each other is consistent with the phase rule of Blair \cite{Am65}.

\section{Conclusions}
We have presented a  predictive model of nucleon-nucleus scattering at
intermediate  energies for  which  nuclear structure  plays a  central
part.  As  such  it   is  most  appropriate  for  eliciting  structure
information of exotic nuclei from analyses of scattering of beams from
hydrogen.  Such  analyses  are  complementary to  electron  scattering
which, given the absence of electron scattering data, makes scattering
from hydrogen the best  (current) means of understanding exotic nuclei
at a microscopic level.

The model has been tested by analyses of proton scattering data across
the mass range and for various  energies. For light nuclei, use of the
shell model with appropriate choices of single particle wave functions
allows for analyses  of data out to the drip  lines. Extension of such
analyses to heavy nuclei is achieved by use of the Skyrme-Hartree-Fock
and RPA models.  In all cases, the underlying  proton scattering model
is  consistent;  analyses  is  predicated  on  a  credible  choice  of
structure model for the target.

It is hoped that facilities utilising exotic beams continue scattering
experiments  with  hydrogen as  the  target,  to  allow for  a  deeper
understanding of the structures of  nuclei out to the drip lines. Such
experiments are not restricted to intermediate energies, for which the
Melbourne  force is appropriate.  With the  advent of  a Multi-Channel
Algebraic  Scattering   theory  (see  \cite{Am05},   this  conference)
analyses of data from low-energy  facilities may be done with the same
predictive power.

\bibliographystyle{aipproc}   % if natbib is available

\bibliography{KarataglidisS}

\end{document}